\documentclass[aps,prl,reprint,superscriptaddress,showpacs,longbibliography,footnoteinbib]{revtex4-2}

\usepackage{physics}

\usepackage[english]{babel}
\usepackage[utf8]{inputenc}
\usepackage[T1]{fontenc}
\usepackage{mathrsfs}
\usepackage{bbm}
\usepackage{bm}
\usepackage{enumerate}
\usepackage{graphicx}
\usepackage[dvipsnames]{xcolor}
\usepackage{float}
\usepackage{subdepth}
\usepackage{fnpct}
\usepackage{booktabs}
\usepackage{multirow}
\usepackage[math]{cellspace}

\usepackage{lipsum}
\usepackage{comment}

\usepackage{lipsum}
\usepackage{comment}

\PassOptionsToPackage{normalem}{ulem}
\usepackage{ulem}
\usepackage[unicode=true,
 bookmarks=false,
 breaklinks=false,pdfborder={0 0 1},backref=false,colorlinks=true]
 {hyperref}

\newcommand\eref[1]{Eq.~\ref{#1}}
\newcommand\fref[1]{Fig.~\ref{#1}}

\newcounter{pr}

\makeatletter

\providecolor{lyxadded}{rgb}{0,0,1}
\providecolor{lyxdeleted}{rgb}{1,0,0}

\DeclareRobustCommand{\lyxsout}[1]{\ifx\\#1\else\sout{#1}\fi}

\usepackage{units}\usepackage{wasysym}

\definecolor{orange}{rgb}{0.50, 0.20, 0.0}

\usepackage{bm}
\usepackage{braket}

\makeatother

\begin{document}
\noindent\begin{minipage}[t]{1\columnwidth}%
\global\long\def\ket#1{\left| #1\right\rangle }%

\global\long\def\bra#1{\left\langle #1 \right|}%

\global\long\def\kket#1{\left\Vert #1\right\rangle }%

\global\long\def\bbra#1{\left\langle #1\right\Vert }%

\global\long\def\braket#1#2{\left\langle #1\right. \left| #2 \right\rangle }%

\global\long\def\bbrakket#1#2{\left\langle #1\right. \left\Vert #2\right\rangle }%

\global\long\def\av#1{\left\langle #1 \right\rangle }%

\global\long\def\tr{\text{tr}}%

\global\long\def\Tr{\text{Tr}}%

\global\long\def\pd{\partial}%

\global\long\def\im{\text{Im}}%

\global\long\def\re{\text{Re}}%

\global\long\def\sgn{\text{sgn}}%

\global\long\def\Det{\text{Det}}%

\global\long\def\abs#1{\left|#1\right|}%

\global\long\def\up{\uparrow}%

\global\long\def\down{\downarrow}%

\global\long\def\vc#1{\mathbf{#1}}%

\global\long\def\bs#1{\boldsymbol{#1}}%

\global\long\def\t#1{\text{#1}}%

\global\long\def\ii{\mathrm{i}}%

\end{minipage}
\title{
Mean-field approach to Mid-spectrum Eigenstates of long-range interacting Quantum Systems
}
\author{Bojan \v{Z}unkovi\v{c}}
\email{bojan.zunkovic@fri.uni-lj.si}

\affiliation{
University of Ljubljana,  
Faculty of Computer and Information Science, Ljubljana, Slovenia
}
\author{Pedro Ribeiro}
\email{ribeiro.pedro@tecnico.ulisboa.pt}

\affiliation{CeFEMA, Instituto Superior Técnico, Universidade de Lisboa, Av. Rovisco
Pais, 1049-001 Lisboa, Portugal}
\affiliation{Beijing Computational Science Research Center, Beijing 100193, China}
\begin{abstract}
We study the equilibrium properties of the spin-$1/2$ XY chain with an infinite-range transverse interaction. At zero temperature, competition between the XY- and the $z$-ordered phases induced by the infinite-range interactions gives rise to a first-order transition upon increasing the transverse coupling. We show that the two gapless points of the XY model behave in fundamentally different ways: isotropic spin chains experience a first-order transition at finite coupling; maximal anisotropic chains overcome a non-algebraic phase transition at zero coupling strength. The phase diagram depicts a first-order reentrant transition that turns second-order along a tricritical line separating a paramagnetic phase from an ordered one at finite temperature. The mean-field approach captures the local properties of the eigenstates and reveals the appearance of a magnetization gap in the spectrum. Global properties, e.g., entanglement entropy, are well approximated only at spectral boundaries. The mean entanglement entropy and the level-spacing ratio deviate from the Gaussian results, revealing the interacting nature of the problem. 
\end{abstract}
\maketitle

\section{Introduction}
The presence of long-range interactions yields quantum many-body systems with rather universal properties. Interactions between distant degrees of freedom decrease the role of fluctuations and render the effective dimensionality larger than the upper critical dimension. Thus, the resulting equilibrium phases and phase transitions are usually well captured by mean-filed methods~\cite{defenu2023long}, which considerably extends the theoretical predictive power for these systems compared to short-range models. As ground states and low-lying excitations are well reproduced, mean-field methods can be reliably utilized to obtain low-energy properties and capture the dynamics, at least for short and intermediate times~\cite{lerose2019impact}. Considering they provide a good approximation of the dynamics, mean-field methods must capture some features of higher excited states. Therefore, it is natural to inquire if it is possible to investigate the nature of highly excited states living in the middle of the spectrum utilizing mean-filed approximations. However, the mean-field approximation typically maps a system into some non-interacting model with self-consistently determined parameters. Eigenstates of such models, namely their entanglement properties, are expected to differ considerably from their non-interacting counterparts. In this work, we give some steps to resolve this apparent paradox by extending the mean-field approach to mid-spectrum eigenstates of a family of long-range interacting spin chain models. A similar problem has been discussed in \cite{den1976systems} and more recently in~\cite{granet2023exact} with the coherent state approach.

One of the most studied spin chain models is the long-range Ising model with two exactly solvable limits. In the nearest neighbor interaction limit, we solve the model by the Jordan-Wigner transformation. The opposite, infinite-range interaction limit, has a permutation symmetry leading to an efficient description in the Dicke basis~\cite{defenu2023long}. At finite interaction strengths, we find a transition between the long-range behavior in the (classical) Ising universality class~\cite{defenu2015fixed} and the short-range universality of the Kitaev chain. Another important family of spin-chain models is the long-range XXZ spin chain integrable in the nearest-neighbor limit. In this limit, standard bosonization techniques lead to an effective sine-Gordon description~\cite{defenu2023long} with the effective charge equal to one. In the long-range interaction regime, the numerical DMRG simulations revealed a larger effective charge~\cite{maghrebi2017continuous}. In \cite{igloi2018quantum}, the XX spin chain model with global transverse anisotropic interaction has been considered. An exact zero-temperature phase diagram has been calculated with two ordered phases (ferromagnetic and antiferromagnetic) and an XY-phase having quasi-longrange order. 

Long-range spin models are increasingly studied also out of equilibrium~\cite{defenu2023out} displaying several unique phenomena, e.g., time-translation symmetry breaking~\cite{sacha2017time,zhang2017observation,choi2017observation,rovny2018observation}, super-luminal information spreading~\cite{richerme2014non,mottl2012roton,jurcevic2014quasiparticle,hauke2013spread,pappalardi2018scrambling}, dynamical phase transition~\cite{sciolla2013quantum,piccitto2019dynamical,vzunkovivc2018dynamical,vzunkovivc2016dynamical}.

Despite increased interest in models with long- and short-range interactions, they are not yet fully understood, with only a few exact analytic results. In this regard, we study the XY spin-1/2 chain with infinite-range transverse interactions. We solve this model with a mean-field approach and provide several analytic and numeric results that are relevant for similar systems with short-long-range interaction dichotomy realized in cavity QED with Quantum Gases \cite{igloi2018quantum, mivehvar2021cavity}. Besides interesting equilibrium properties (e.g., non-analytic critical point and a reentrant phase transition), we show that the mean-field approximation satisfactorily describes the model's spectral and local eigenstate properties. 

While writing the manuscript, we became aware of~\cite{granet2023exact}, where a similar analytical study based on coherent states has been performed for the Ising model. We expand on \cite{granet2023exact} by considering a more general class of models and discussing entanglement properties and level spacing ratios. We also contrast the local with the global properties of mean-field eigenstates. 

We introduce the model and methods in Section~\ref{sec: model}. Then, in Section~\ref{sec: quantum PD}, we study the phase diagram and derive several exact analytical results, including a non-algebraic behavior of the order parameter close to the phase transition and a reentrant phase transition. In Section~\ref{sec: specter}, we study the spectral and eigenstate properties of the model. We conclude in Section~\ref{sec: conclusion}.

\section{Model \label{sec: model}}
We consider an XY spin-$1/2$ chain with long-range couplings along the transverse direction, given by the Hamiltonian 
\begin{align}
H & =-J\sum_{i=1}^{L-1}\left(\cos\eta\,\sigma_{i}^{x}\sigma_{i+1}^{x}+\sin\eta\,\sigma_{i}^{y}\sigma_{i+1}^{y}\right)\nonumber \\
 & -\frac{g}{2L}\sum_{i,j=1}^{L}\sigma_{i}^{z}\sigma_{j}^{z}\label{eq: Hamiltonian}
\end{align}
where $i=1,...,L$ , $\eta\in\left[0,2\pi\right[$ parametrizes the
XY anisotropy, $g$ the transverse coupling, $J$ the XY-coupling
(set to unity in the following), and $\sigma^{\rm x,y,z}_j$ are the Pauli matrices acting on site $j$.

Before turning to the mean-field treatment employed in the rest of the paper, let us discuss the limits of weak and strong infinite-range coupling.  

In the absence of the infinite-range coupling (i.e., $g=0$), the system reduces to a well-studied XY model exactly solvable by the Jordan-Wigner mapping to an equivalent free fermionic system~\cite{lieb1961two}. In this case, we have a ferromagnetic ground-state for $\eta\in\left]-\pi/4,3\pi/4\right[$. The ferromagnetic ground state is characterized by a non-vanishing order parameter $\phi_{x}=\frac{1}{L}\sum_{i=1}^{L}\av{\sigma_{i}^{x}}$ (for $\eta\in\left]-\pi/4,\pi/4\right[$) and $\phi_{y}=\frac{1}{L}\sum_{i=1}^{L}\av{\sigma_{i}^{y}}$ (for $\eta\in\left]\pi/4,3\pi/4\right[$). The points $\eta=\pm\pi/4$ correspond to the phase transitions between the two ordered phases in which the ground state is gapless with power-law correlation functions. In the antiferromagnetic case arising outside the interval $\eta\in\left]-\pi/4,3\pi/4\right[$, we map the system to the ferromagnetic one by the transformation $\sigma_{i}^{\alpha=x,y}\to\left(-1\right)^{i}\sigma_{i}^{\alpha=x,y}$. Away from the zero-temperature limit, the low dimensionality of the system precludes long-range order, and only the paramagnetic phase can be found.

In the limit $g/J\gg1$, the transverse, infinite-range coupling dominates, and the model is exactly solvable in the large-$L$ limit where a mean-field approach becomes exact. The order parameter, $\phi_{z}=\frac{1}{L}\sum_{i=1}^{L}\av{\sigma_{i}^{z}}$, is non-zero for $g>0$ from $T=0$ up to a critical temperature $T_{c}$. At this second-order critical point, physical observables obey scaling laws with mean-field scaling exponents~\cite{sachdev2007quantum}.

We now consider the thermodynamic limit, $L\to\infty$ limit, and follow a mean-field approximation in the presence of the longitudinal coupling by employing the decoupling ansatz $\sigma_{i}^{z}\sigma_{i}^{z}\to\av{\sigma_{i}^{z}}\sigma_{i}^{z}+\sigma_{i}^{z}\av{\sigma_{i}^{z}}-\av{\sigma_{i}^{z}}\av{\sigma_{i}^{z}}$ in Eq.\eqref{eq: Hamiltonian}. The resulting mean-field Hamiltonian reduces to that of an XY chain in a transverse field 
\begin{align}
H_{\text{MF}}= & -J\sum_{i=1}^{L-1}\left(\cos\eta\,\sigma_{i}^{x}\sigma_{i+1}^{x}+\sin\eta\,\sigma_{i}^{y}\sigma_{i+1}^{y}\right)\nonumber \\
 & -h\sum_{i=1}^{L}\sigma_{i}^{z}+\frac{1}{2g}h^{2},\label{eq: mean-field H}
\end{align}
together with the self-consistency condition 
\begin{align}
h= & \frac{g}{L}\sum_{i=1}^{L}\av{\sigma_{i}^{z}}.\label{eq: selfcond}
\end{align}
$H_{\text{MF}}$ also admits an exact solution by the Jordan-Wigner mapping to an equivalent quadratic fermionic system. Therefore, its free-energy per unit length $f_{\t{MF}}=-\frac{1}{\beta L}\ln\tr\left(e^{-\beta H_{\t{MF}}}\right)$ is given by 
\begin{align*}
f_{\t{MF}} & =\frac{1}{2g}h^{2}-\frac{1}{\beta}\int_{-\pi}^{\pi}\frac{dk}{2\pi}\ln\left[2\cosh\left(\beta\omega_{k}\right)\right]
\end{align*}
with the dispersion relation~\cite{lieb1961two} {\small{}
\begin{align*}
\omega_{k} & =\sqrt{\left[h+J\left(\cos\eta+\sin\eta\right)\cos k\right]^{2}+\left(1-\sin2\eta\right)\left(J\sin k\right)^{2}}.
\end{align*}
}{\small\par}

\section{Equilibrium Phase diagram}
\label{sec: phase diagram}

In the following, we obtain the phase diagram of the model at zero and finite temperature by minimizing the free energy density with respect to the order parameter, $\pd_{h}f_{\t{MF}}=0$, or equivalently by solving the self-consistency condition in Eq. \eqref{eq: selfcond}, and analyzing the nature of the ensuing solutions. For simplicity, we limit our phase diagram analysis to the region $\eta\in\left]-\pi/4,\pi/4\right[$. The interval $\eta\in\left]\pi/4,3\pi/4\right[$ can be obtained by a permutation $x\leftrightarrow y$ and thus has similar properties.

\subsection{Zero Temperature \label{sec: quantum PD}}

\begin{figure}[!h]
\centering
\includegraphics[width=0.85\columnwidth]{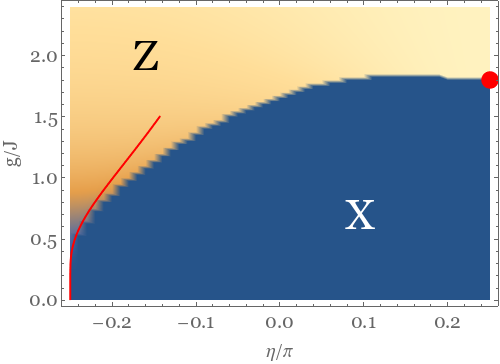}
\includegraphics[width=0.1\columnwidth]{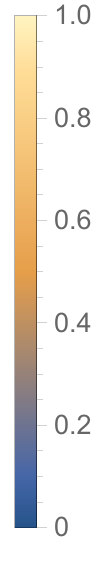}

\includegraphics[width=1\columnwidth]{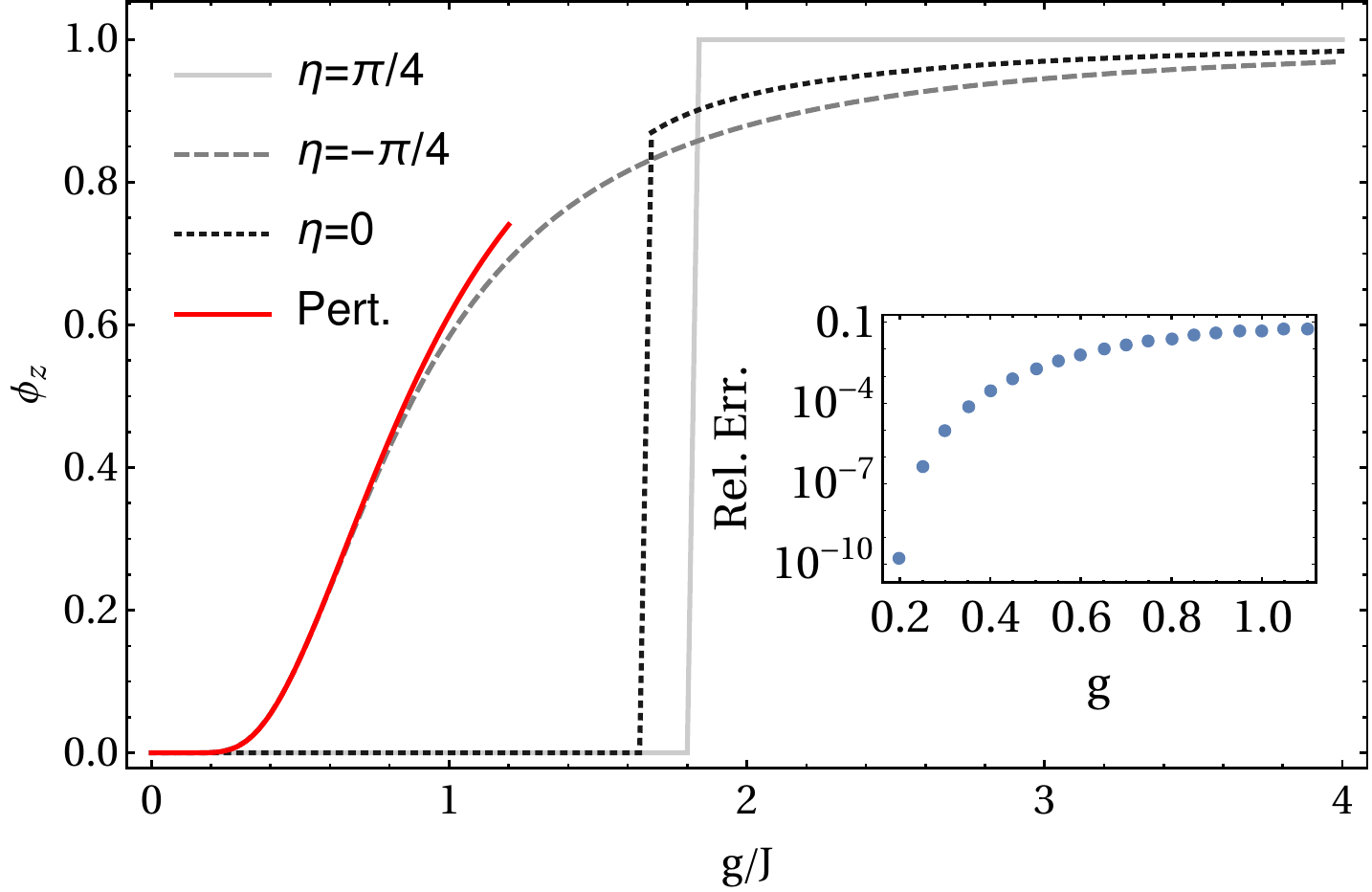}    
\caption{\label{fig: quantum_phase_diagram} a) 
Zero temperature phase diagram. The transition between the \textit{X} and \textit{Z} phases is first order (see subfigure b), except at the point $\eta=-\pi/4$. The red dot represents the analytical result for the XX model $g_{\rm critical}/J=4\sqrt{2}/\pi$ derived in Appendix \eqref{sec: Further-analysis-of-Phase Diagram}. The red line is the approximate phase transition line close to $\eta=-\pi/4$ given by \eref{eq: critical angle}. \textbf{b)} Order parameter of the Z phase, $\phi_{z}$, as a function of the infinite-range interaction strength, $g$, for different values of $\eta$. The discontinuity of $\phi_{z}$ at the phase transition ranges from unity, at $\eta=\pi/4$, to zero at $\eta=-\pi/4$. In the vicinity of the critical point, $(\eta,g)=(-\pi/4,0)$, the order parameter is captured by a perturbative result \eref{eq: 3quat z-1} (red line). The relative error of the \eref{eq: 3quat z-1} is shown in the inset of the figure b).}
\end{figure}

The zero-temperature phase diagram is depicted in~\fref{fig: quantum_phase_diagram}-a) It features two ordered phases, labeled $X$ and $Z$, characterized by a non-vanishing order parameter $\phi_{x}$ or $\phi_{z}$, and separated by a phase transition line. Fig.\ref{fig: quantum_phase_diagram}-b) depicts the order parameter of the $Z$ phase, $\phi_{z}$, as a function of $g$, across the transition. As the two ordered phases break distinct symmetries, the phase transition is generically discontinuous (first-order). The largest jump in the order parameter is observed at the right boundary, i.e., at $\eta=\pi/4$, where the order parameter jumps to one at the critical infinite-range interaction strength $g_{\pi/4}^*=4\sqrt{2}/\pi$ (see Appendix \ref{sec: Further-analysis-of-Phase Diagram}). At the left boundary, $\eta=-\pi/4$, the discontinuity in $\phi_{z}$ vanishes at a continuous critical point $(\eta,g)=(-\pi/4,0)$, for which a perturbative approach (see Appendix \ref{sec: Further-analysis-of-Phase Diagram} ) shows that the non-analyticity of the order parameter is given by
\begin{equation}
\phi_{z}\left(g,\eta=-\frac{\pi}{4}\right)=\frac{4\sqrt{2}e^{-\frac{\pi}{\sqrt{2}g}}}{g}.\label{eq: 3quat z-1}
\end{equation}
Denoting the discontinuous phase transition line as $g^{*}\left(\eta\right)$, near $\eta=-\pi/4$, one finds 
\begin{equation}
\eta=-\pi/4+4e^{-\left(1+\frac{\pi}{\sqrt{2}g^{*}\left(\eta\right)}\right)}.\label{eq: critical angle}
\end{equation}
We refer the reader to Appendix \eqref{sec: Further-analysis-of-Phase Diagram} for the details on the perturbative approach and further analysis of this transition point. 

\subsection{Finite Temperature}
\label{sec: T PD}
\begin{figure}[!!htb]
\centering

\includegraphics[width=0.95\columnwidth]{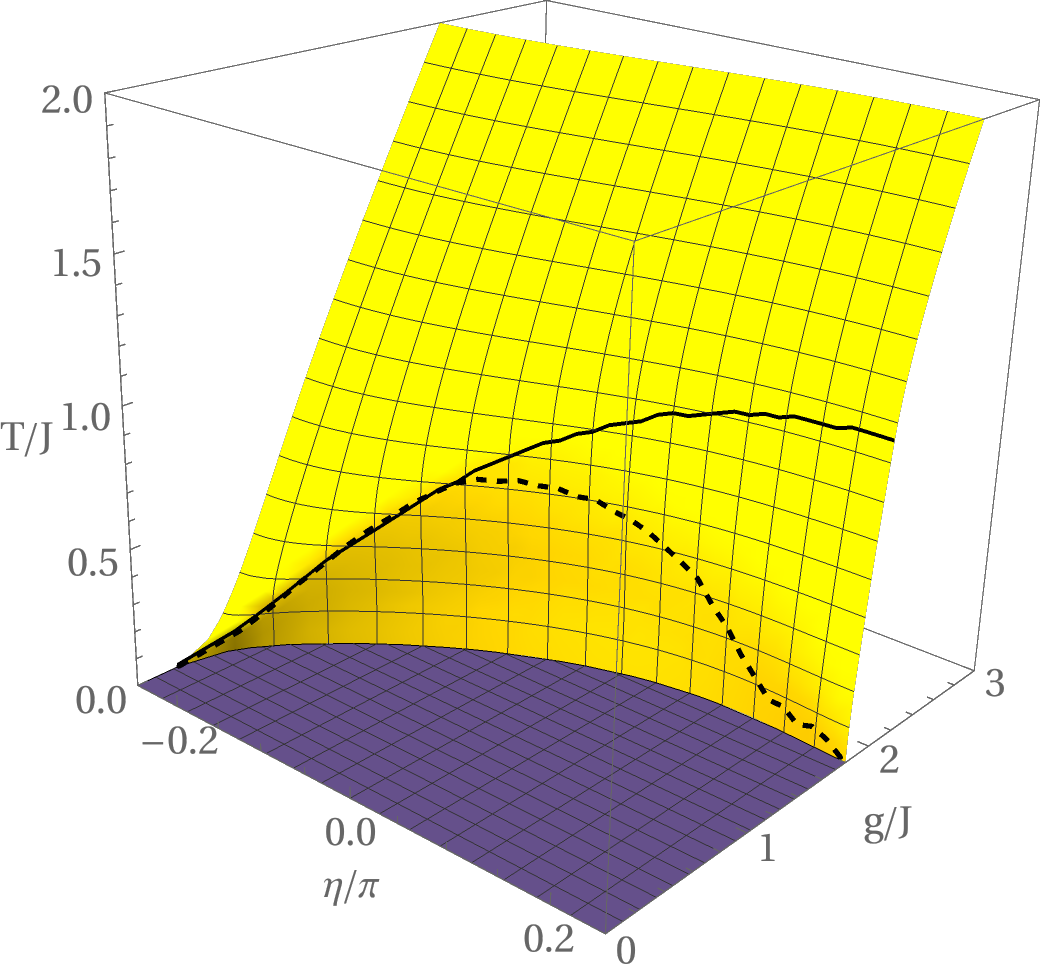}


\includegraphics[width=0.98\columnwidth]{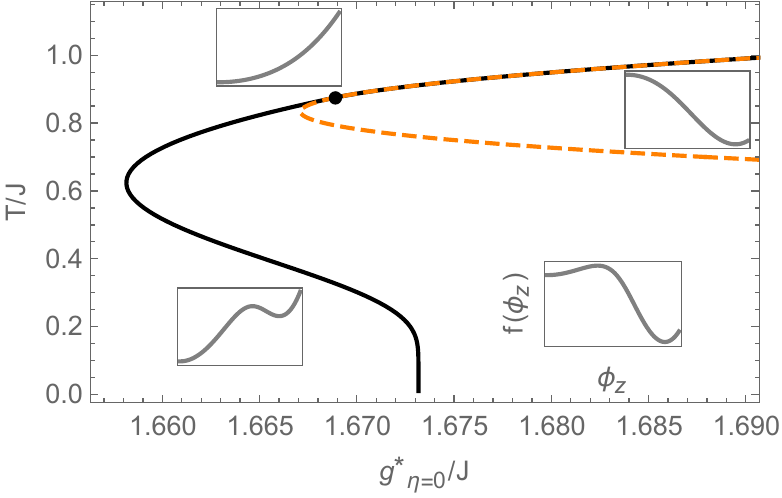}

\caption{a) The critical surface of the finite temperature phase transition is shown in yellow. The X phase at  $T=0$ is depicted in dark blue. The solid line represents the tricritical line. On the surface above this line (higher temperature or larger $g$), the phase transition is second order, while below it is discontinuous. The dashed line shows the smallest critical $g$ at fixed $\eta$ that delimits the boundary of the reentrant phase transition. b) reentrant phase transition in the Ising model $\eta=0$ (black line). The dashed line represents the points where the paramagnetic solution becomes unstable. The dot is the tricritical point. The insets show representative free energy shapes in various regions of the reentrant phase transition.}
\label{fig: T_pt}
\end{figure}

Fig.\eqref{fig: T_pt} shows the finite temperature phase diagram. The $Z$-ordered phase extends to $T>0$ stabilized by the long-range $z$-interactions. In contrast, the short-range nature of the $x$-interaction terms renders the $X$ phase unstable at finite temperature, and thus, for $T>0$, $\phi_{x}=0$. Nonetheless, the discontinuous phase transition between the $Z$ phase and the finite $T$ disordered state persists at low temperatures within the $\eta-g$ plane. At high temperatures, the magnetic/non-magnetic transition becomes second-order. The passage from a discontinuous to a continuous transition occurs at the tricritical line depicted as a full black line in Fig.\eqref{fig: T_pt}-a) and as a dashed orange line in Fig.\eqref{fig: T_pt}-b).

For some regions of the $\eta-g$ diagram, a reentrant phase transition is observed as a function of temperature, see Fig.\eqref{fig: T_pt}-b). Here, for some $(\eta,g)$ points corresponding to a disordered low-temperature state, there is a discontinuous transition to the ordered Z-phase upon increasing the temperature. Further increasing $T$ makes the system transition again to the disordered non-magnetic phase. The dashed line in \fref{fig: T_pt}-a) depicts the boundary $(\eta,g,T)$ of the reentrant transition, i.e. the smallest critical $g$ at fixed $\eta$. Fig.~\ref{fig: T_pt}~b shows the reentrant phase transition in the Ising case ($\eta=0$).

In the limit of large coupling, $J/g\ll1$, the critical temperature, and the critical exponents may be computed perturbatively, yielding 
\begin{align}
T_{J/g\ll1}^{*}/g & \approx1+O((J/g)^{2}).
\end{align}
Furthermore, by expanding the free energy around the critical fixed point, we obtain $\phi_{z}(J/g\ll1)\approx\sqrt{3}|1-\Delta T|^{1/2}$, corresponding to the expected mean-field exponent $\beta=1/2$. This value of $\beta$ arises for all continuous finite-temperature transitions away from the tricritical line (solid balk line in Fig \eqref{fig: T_pt}-(a)), specifying the passage to a discontinuous transition. Along the tricritical line, the exponent retains its mean-field nature now with $\beta=1/4$. These results are obtained by numerically minimizing the free energy and by perturbative calculations in the Ising limit, $\eta=0$. We relegate the details to Appendix \eqref{sec: Further-analysis-of-Phase Diagram}. 

\section{Spectral and eigenstate properties \label{sec: specter} }

In this section, we obtain a variational approximation for the eigenstates of the Hamiltonian of Eq.\eqref{eq: Hamiltonian} in the spirit of mean-field and compare the spectral and eigenstate properties to those obtained by exact diagonalization. 

\subsection{Mean-field equations of motion }
To arrive at the mean-field equations of motion, we first rewrite the mean-field model in the pseudospin representation and then expand the propagator in terms of $1/L$.

\subsubsection{ Pseudospin representation} 
We start by considering the Jordan-Wigner transformed Hamiltonian, implemented by operators $a_{r}^{\dagger}=e^{\ii\pi\sum_{r'=0}^{r-1}\sigma_{r'}^{+}\sigma_{r'}^{-}}\sigma_{r}^{+}$, that obey fermionic commutation relations, $\left[a_{r}, a_{r'}^{\dagger}\right]=\delta_{rr'}$. We transform the $z-$coupling terms in~\eref{eq: Hamiltonian} using $\sigma_{r}^{z}=2a_{r}^{\dagger}a_{r}-1$ and the $XY$ component as
\begin{align*}
H_{XY}= & -J\sum_{r=1}^{L}\left[\left(\cos\eta+\sin\eta\right)\left(a_{r+1}^{\dagger}a_{r}+a_{r}^{\dagger}a_{r+1}\right)\right.\\
 & \left.+\left(\sin\eta-\cos\eta\right)\left(a_{r}a_{r+1}+a_{r+1}^{\dagger}a_{r}^{\dagger}\right)\right].
\end{align*}
The fermionic, mean-field Hamiltonian in the momentum basis, $a_{q}^{\dagger}=\frac{1}{\sqrt{L}}\sum_{r}e^{-\ii qr}a_{r}^{\dagger}$, is given by
\begin{align}
H_{\t{MF}}(t)& =H_{XY}-h(t)\frac{1}{L}\sum_{q}\left(2a_{q}^{\dagger}a_{q}-1\right)+\frac{L}{2g}h^{2},\\ \nonumber
H_{\rm MF}&=\\ \nonumber
h(t)&=\frac{g}{L}\sum_q\left\langle2a^\dag_qa_q-1\right\rangle
\end{align}
Such paring Hamiltonian is most conveniently written in terms of Anderson\textquoteright s pseudospin representation $\vec{\tau}_{q}=\psi_{q}^{\dagger}.\vec{\tau}.\psi_{q}$, with $\psi_{q}=\left\{ a_{q},a_{-q}^{\dagger}\right\} ^{T}$ the Nambu spinor and $\vec{\tau}=\left\{ \tau_{x},\tau_{y}, \tau_{z}\right\} ^{T}$ are the Pauli matrices. Note that the pseudospin satisfies the usual commutation relations for angular momentum, $\left[\tau_{q}^{\alpha},\tau_{q'}^{\beta}\right]=2i\delta_{qq'}\varepsilon_{\alpha\beta\gamma}\tau_{q}^{\gamma}$. Using this representation, we have
\begin{align}
H_{\t{MF}}\left(t\right) & =\sum_{q}\vec{B}_{q}\left[h\left(t\right)\right]\cdot\vec{\tau}_{q}\label{eq: Hamiltonian pseudospin}
\end{align}
with
\begin{align}
\vec{B}_{q}\left(h\right) & =\left\{ 0,4\sin(q)cs_-,4\left[\cos(q)cs_++h\right]\right\} ^{T} \\ \nonumber
cs_-&=\cos\eta-\sin\eta,\quad cs_+=\cos\eta+\sin\eta
\end{align}
and
\begin{align}
    h=\frac{g}{L}\sum_q\langle\tau^{\rm z}_q\rangle. \label{eq: self-consistency pseudospin}
\end{align}

\subsubsection{ Propagator}
We now consider the propagator, $K\left(t\right)=\bra{\vec{\phi}'}e^{-\ii Ht}\ket{\vec{\phi}}$ between two Slater determinant states $\ket{\vec{\phi}}$ and $\ket{\vec{\phi}'}$ determined by the original Hamiltonian \eref{eq: Hamiltonian}. Using the identity 
\begin{multline*}
e^{-\ii H\Delta t}\simeq e^{-\ii H_{XY}\Delta t}e^{\ii\frac{g}{2L}\left(\sum_{i}\sigma_{i}^{z}\right)\left(\sum_{j}\sigma_{j}^{z}\right)\Delta t}+O\left(\Delta t^{2}\right),\\
\simeq\int dh \, e^{-\ii\left(H_{XY}+\frac{L}{2g}h^{2}-\frac{h}{L}\left(\sum_{i}\sigma_{i}^{z}\right)\right)\Delta t}+O\left(\Delta t^{2}\right),
\end{multline*}
we write the propagator as 
\begin{align}
K\left(t\right) & =\int Dh\ \bra{\vec{\phi}'}\hat{T}e^{-\ii\int_{0}^{t}dt'H_{\t{MF}}\left(t\right)}\ket{\vec{\phi}}\label{eq: propagator}
\end{align}
where $Dh$ denotes path integration and $\hat{T}$ the time ordering
operator. We evaluate the propagator in Eq.\eqref{eq: propagator} in the $L\to\infty$ limit by a saddle-point expansion, yielding
\begin{align}
h\left(t\right) & =\frac{\bra{\vec{\phi}'}\frac{g}{L}\sum_{i}\sigma_{i}^{z}\hat{T}e^{-\ii\int_{0}^{t}dt'H_{\t{MF}}\left(t\right)}\ket{\vec{\phi}}}{\bra{\vec{\phi}'}\hat{T}e^{-\ii\int_{0}^{t}dt'H_{\t{MF}}\left(t\right)}\ket{\vec{\phi}}}.
\end{align}
We shall assume the $\ket{\vec{\phi}'}$ lays within the mean field trajectory, i.e. $\ket{\vec{\phi}'}=\ket{\vec{\phi}\left(t\right)}=\hat{T}e^{-\ii\int_{0}^{t}dt'H_{\t{MF}}\left(t\right)}\ket{\vec{\phi}}$. Furthermore, defining the pseudo-magnetization vector for the $q$-mode, $\vec{\phi}_{q}\left(t\right)=\bra{\vec{\phi}\left(t\right)}\vec{\tau}_{q}\ket{\vec{\phi}\left(t\right)}$, the saddle-point equation reduces to
\begin{align}
h\left(t\right) & =\frac{g}{L}\sum_{q}\phi_{q}^{z}\left(t\right),\label{eq: h_of_t}
\end{align}
whereas the time evolution of $\vec{\phi}_{q}\left(t\right)$ is given by
\begin{align}
\pd_{t}\vec{\phi}_{q}\left(t\right) & =\ii\bra{\vec{\phi}\left(t\right)}\left[H_{\t{MF}}\left(t\right),\vec{\tau}_{q}\right]\ket{\vec{\phi}\left(t\right)},\nonumber \\
 & =\vec{B}_{q}\left[h\left(t\right)\right]\times\vec{\phi}_{q}\left(t\right).\label{eq: phi_of_t}
\end{align}
Obtained equations Eqs.\eqref{eq: h_of_t} and \eqref{eq: phi_of_t}  can also be viewed as Heisenberg equations concerning the mean-field Hamiltonian given in Eqs. \eqref{eq: Hamiltonian pseudospin} and \eqref{eq: self-consistency pseudospin}. 

\subsection{Approximate Eigenstates}
In this section, we consider stationary states of the equations of motion, $\pd_{t}\vec{\phi}_{q}^{*}=0$, for which
\begin{align} 
0 & =\vec{B}_{q}\left(h^{*}\right)\times\vec{\phi}_{q}^{*},\label{eq: stationarity_1}\\
h^{*} & =\frac{g}{2L}\sum_{q}\phi_{q}^{*z},\label{eq: stationarity_2}
\end{align}
as approximations of the many-body eigenstates of the exact model. In terms of the propagator at the saddle-point level, the stationary condition translates to 
\begin{align*}
\ln K^{*} & =-\ii \bra{\vec{\phi}^{*}}H_{\t{MF}}\ket{\vec{\phi}^{*}}t+O\left(L^{0}\right),
\end{align*}
meaning that $\ket{\vec{\phi}^{*}}$ is an approximate eigenstate of $H$ with an approximate energy 
\begin{align}
e^{*}= & \bra{\vec{\phi}^{*}}H_{\t{MF}}\ket{\vec{\phi}^{*}}.
\end{align}

To sample different eigenstates, we note that there are several ways of satisfying Eqs. \eqref{eq: stationarity_1}-\eqref{eq: stationarity_2} depending on whether the field and the magnetization of mode $q$ are chosen to be parallel or antiparallel $\vec{B}_{q}\cdot\vec{\phi}_{q}^{*}=\xi_{q}\abs{\vec{B}_{q}}\abs{\vec{\phi}_{q}^{*}}$ with $\xi_{q}=\pm1$. Since for a pure state $\abs{\vec{\phi}_{q}^{*}}^{2}=1$, for each of the $2^{L}$ possible choices $\xi=\left\{ \xi_{0}=\pm1,\xi_{2\pi/L}=\pm1,...\right\} $, a solution of the self-consistency condition of Eq.\eqref{eq: stationarity_2} with 
\begin{align} \label{eq: xi_q}
\vec{\phi}_{q}^{*} & =\xi_{q}\frac{\vec{B}_{q}\left(h^{*}\right)}{\abs{\vec{B}_{q}\left(h^{*}\right)}},
\end{align}
solves the static saddle-point condition. The solution with $\xi_{q}=-1$ corresponds to the zero temperature case obtained previously. In this case, there are one or two solutions ($h=0$ and $h\neq0$) depending on the parameters, which gives rise to the quantum phase diagram discussed in the previous sections. We can also find multiple solutions for other configurations of $\xi$. Since there are $2^L$ different configurations with at least one solution, the number of approximate eigenstates obtained by our mean-field approach exceeds the dimension of the Hilbert space. Nevertheless, the mean-field approach can still capture the behavior of \textit{local} observables. On the other hand, it can not describe global properties, such as entanglement entropy. Which spectral and eigenstate properties are correctly described by the mean-field analysis is studied in the following sections.

\subsection{Spectral properties and microcanonical order parameter}

With the eigenstate construction described in the previous section, we now turn to the study of spectral and eigenstate properties. The advantages and drawbacks of the mean-field approach are discussed by contrasting its predictions with exact diagonalization for small-size systems. Subsequently, we utilize the mean-field approach to obtain eigenstate properties for large system sizes. 

We first discuss the spectral structure and the density of states in the $\phi_{\rm z}$-$e$ plane, where $e$ denotes the energy density. In this regard, we need to sample eigenstates from the correct distribution. Sampling approximate eigenstates amounts to choosing an assignment of $\xi_q=\pm1$. However, doing so randomly yields predominantly states in the middle of the spectrum with a roughly equal number of positive and negative $\xi_q$. We overcome this problem by fixing the number of excitations, i.e., positive  $\xi_q$'s, which enables efficient sampling of low- and high-energy states. After sampling a configuration in a given excitation sector, we solve the self-consistency equations \eref{eq: stationarity_2} and retain all solutions. We obtain the density of states by first sampling over sectors with different excitation numbers, $M$, and then summing the sector-resolved densities weighted by the entropic factor $\binom{L}{M}$. 

\begin{figure}[!!htb]
\includegraphics[width=0.475\columnwidth]{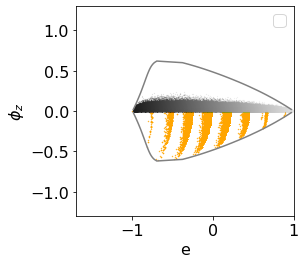}
\includegraphics[width=0.51\columnwidth]{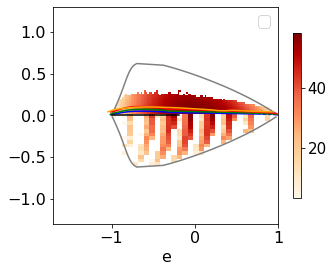}
\includegraphics[width=0.475\columnwidth]{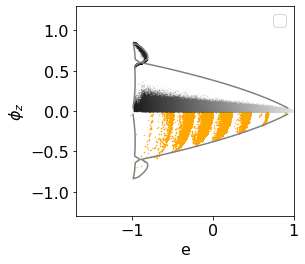}
\includegraphics[width=0.51\columnwidth]{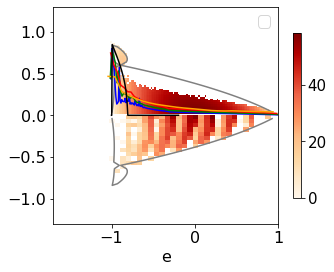}
\includegraphics[width=0.475\columnwidth]{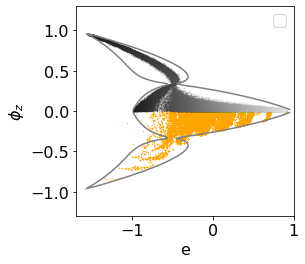}
\includegraphics[width=0.51\columnwidth]{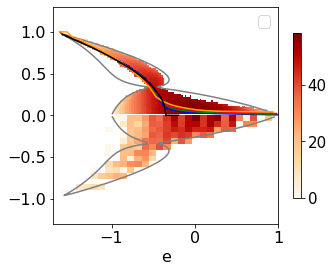}
\caption{
Eigenstate energy versus $z$-axes magnetization obtained for $\eta=0$ and $g=1., 1.67$, and $3.$ (from top to bottom). 
\textit{Left panels:} Results for eigenstates obtained by exact diagonalization with $L=17$ are given in the lower half-plane (yellow dots). Mean-field prediction of the eigenstates is plotted in the upper half-plane (gray dots). The grayscale corresponds to the number of excitations (increasing from dark to bright), i.e., modes with $\xi_q = 1$ in Eq.(\ref{eq: xi_q}), used to solve the self-consistency condition Eqs. (\ref{eq: stationarity_1}-\ref{eq: stationarity_2}). The black lines are the mean-field spectral boundaries obtained in the thermodynamic limit. \textit{Right panels:} Approximate density of states. The color scale corresponds to the logarithm of the normalized number of states. Colored lines represent the mean-field microcanonical ensemble prediction for system sizes $L=50$ (orange), $L=100$ (red), $L=200$ (green), and $L=400$ (blue). The black line shows the canonical prediction in the thermodynamic limit ($L\to\infty$) presented in Section \ref{sec: phase diagram}. Model parameters: (\textit{top
-- paramagnet}) $g=1,\eta=0$, (\textit{middle -- reentrant phase transition}) $g=1.67,\eta=0$, and (\textit{bottom - second order phase transition}) $g=3,\eta=0$.
}
\label{fig: ez specter}
\end{figure}

The \fref{fig: ez specter} depicts the energies resolved by the value of the order parameter, $\phi_{\rm z}$, for different model parameters. We compare the exact diagonalization results for a small system size, $L=17$, with the mean-field results for a large system size, $L=200$. The mean-field results were obtained by sampling $10^5$ random configurations for each excitation sector $M=1,\ldots L$. States with positive and negative $\phi_{\rm z}$ have the same energy due to the spin-flip symmetry. Therefore, we show the mean-field results on the upper half-plane and the exact diagonalization results on the lower half-plane. 

The left panels of \fref{fig: ez specter} show $(e,\phi_z)$ pairs for sampled eigenstates. We observe well-separated excitation sectors at small system sizes (exact diagonalization), which merge in the thermodynamic limit. As expected, lower energies obtained in the mean-field case have a smaller number of excitations, which we demonstrate by the brightness of the $(e,\phi_z)$ points  -- black corresponds to a small and light gray to a large number of excitations.

The right panels of \fref{fig: ez specter} depict the density of states in the $(e,\phi_z)$ plane, which drops exponentially towards the boundary of the spectrum. Overall, the spectral shape is well reproduced by the mean-field approach in all regimes. We also show the mean-field prediction of the spectral edge (black lines) obtained in the thermodynamic limit by extremazing the energy for each magnetization sector (see Appendix~\ref{app: spectral boundaries} for details). 

Further, we compare the finite-size mean-field microcanonical order parameter with its canonical value in the thermodynamic limit. In \fref{fig: ez specter} right panels, the colored lines represent microcanonical predictions for different systems sizes, $L$, and the black line is the thermodynamic canonical result.  The mean-field canonical calculations are well-matched with the microcanonical results in the paramagnetic and ferromagnetic regions. Since the reentrant phase transition region is very narrow, we observe many fluctuations in these cases. Still, the qualitative behavior predicted by the mean-field canonical treatment discussed in Section \ref{sec: phase diagram} is well reproduced by the mean-field microcanonical results. We observe a square-root system size convergence of the micro-canonical predictions to the infinite system size canonical result~(see Appendix \ref{app: microcanonical convergence} for details).

The agreement between the mean-field canonical and the mean-field microcanonical descriptions is good even in regimes with three or fivefold degeneracy of the solutions. We showcase the validity of degenerate solutions in the ferromagnetic case corresponding to the bottom panels of \fref{fig: ez specter}. 

\begin{figure}[!htb]
   \includegraphics[width=0.95\columnwidth]{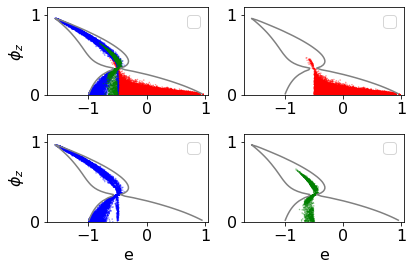}
    \caption{
    The number of solutions for the same configuration of excitations $\xi$: red (one solution), blue (three solutions), green (five solutions). The top left panel shows all solutions and the remaining three show only solutions with a certain degeneracy. Parameters: $g=3$, $\eta=0$, $L=200$.
}
\label{fig:  number of solutions}
\end{figure}

In \fref{fig:  number of solutions}, we disentangle one-, three-, or five-fold degenerate solutions of the self-consistency equation. We observe that the ferromagnetic phase is composed of states with at least three-fold degeneracy and still matches the mean-field canonical prediction and qualitatively agrees with the exact spectrum at finite system size $L=17$ (shown in \fref{fig: ez specter}). We explain the robustness of microcanonical properties with the entropic factor. Distinct branches of degenerate solutions have very different energies at the same filling. Hence, the largest energy branches at some filling have the same energy as the lowest energy solutions with a higher filling. At the same time, their entropic factors differ exponentially due to the difference in filling. Therefore, only one solution is thermodynamically relevant.

\begin{figure}[!h]
\includegraphics[width=0.45\textwidth]{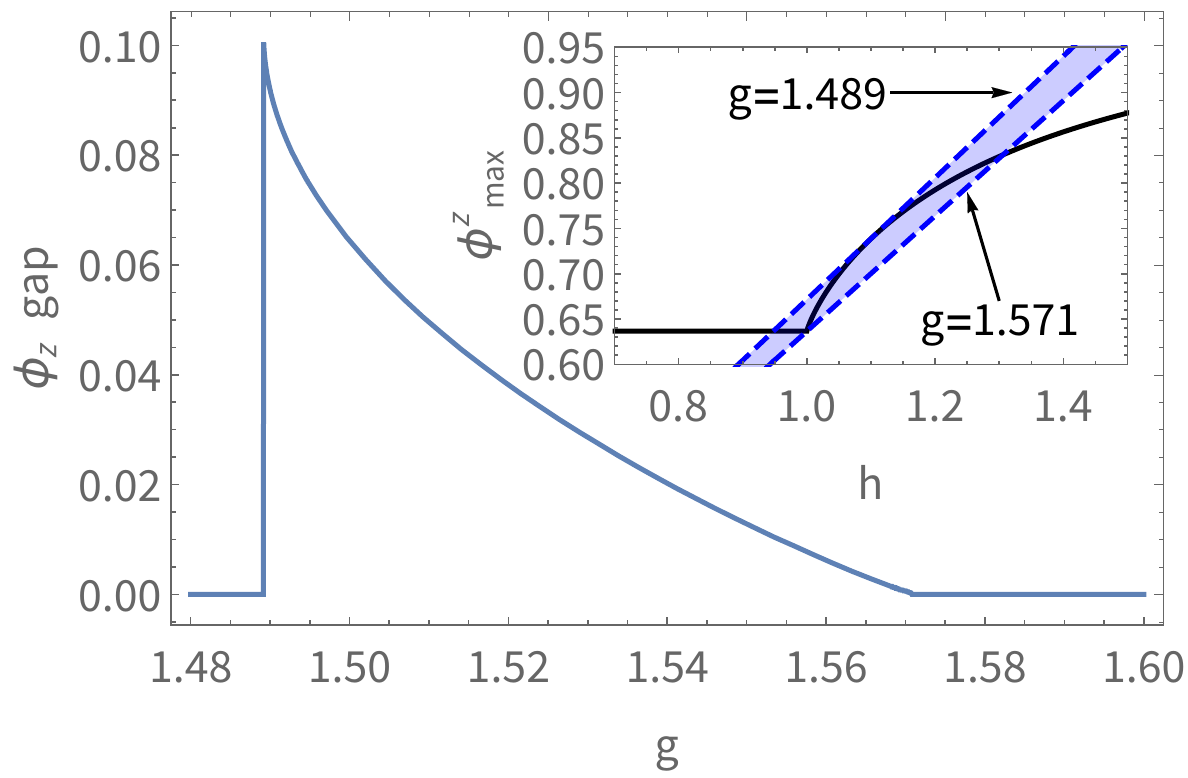}\\[0.6cm]
\includegraphics[width=0.45\textwidth]{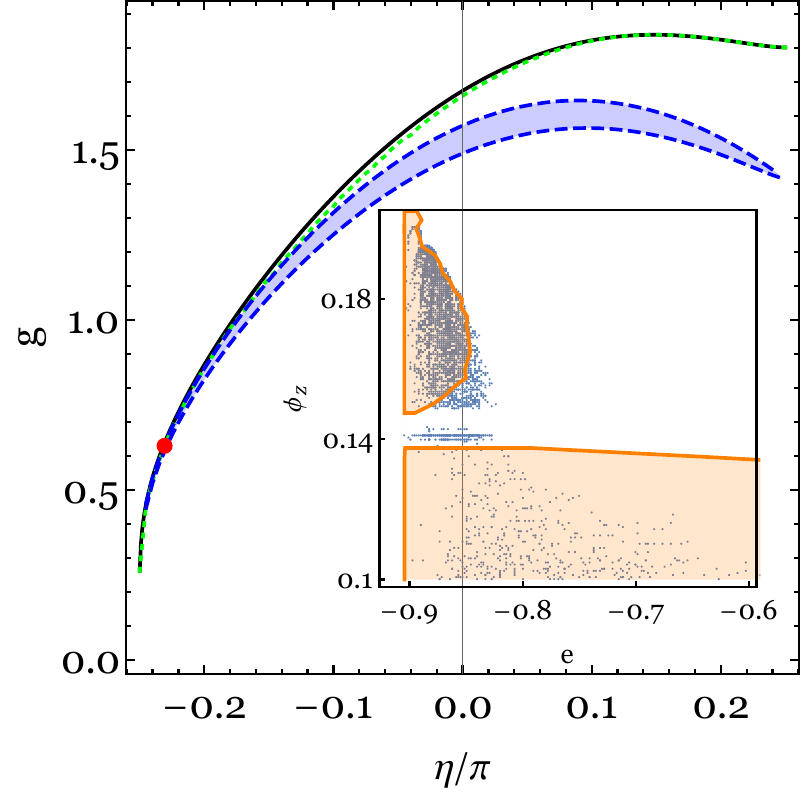}
\caption{(\textit{Top}) Eigenstate magnetization ($\phi_z$) gap in the case of the Ising model, $\eta=0$. The inset demonstrates allowed solutions. The dashed lines represent the self-consistency condition $h/g$ at the boundaries $g=1.489$ and $g=1.571$. The light blue region denotes a region with a finite magnetization gap. The full line corresponds to the maximum allowed magnetization $s_{max}^z(h)$. Only states with $h/g\leq s_{max}^z(h)$ are allowed in the thermodynamic. (\textit{Bottom}) The black line corresponds to the quantum phase transition. The green dotted line denotes the smallest critical field $g$ at any temperature. Hence, the region between the black and the green line corresponds to a parameter region with a reentrant phase transition. The dashed blue lines determine the boundaries of the parameter region (light blue), where we observe a magnetization gap in the spectrum. The inset depicts the comparison between the mean-field spectral boundaries in the thermodynamic limit and the exact finite-size spectrum ($L=17$) in the region of the magnetization gap. We denote the parameters of the inset by the red dot in the main figure.}
\label{fig:  magnetization gap}
\end{figure}

Besides the general form of the spectrum and the mean-field order parameter, the mean-field eigenstates also capture fine spectral features. For example, the spectrum splits into two bands with a finite transverse magnetization gap in a specific parameter region. We numerically calculate the thermodynamic mean-field prediction shown in \fref{fig:  magnetization gap}. First, we calculate the maximum eigenstate magnetization of the XY model with a given transverse field $h$. We compare this upper bound to the self-consistency condition $\phi_{\rm z}=h/g$ corresponding to a line with a coefficient $1/g$. The allowed eigenstate magnetization lies on the self-consistency line below the maximum magnetization threshold. By varying $g$, we find a region where the self-consistency condition line intersects the maximum allowed magnetization threshold three times, giving rise to two separate magnetization bands (see \fref{fig:  magnetization gap}a). The parameter region with two magnetization bands is present at all interaction angles $\eta$ as shown in \fref{fig:  magnetization gap}b. We confirm the appearance of the spectral magnetization gap by finite-size exact diagonalization calculations (inset of \fref{fig:  magnetization gap}b).

\subsection{Entanglement entropy and level spacing ratios}
We discussed several interesting spectral features captured by the mean-field calculations in the previous section. However, due to the over-counting of solutions, we expect that the mean-field treatment will not reproduce the global properties of the exact model. In this regard, we investigate the entanglement distribution over the eigenstates and the distribution of level spacing ratios. In both cases, the self-consistent mean-field solution gives predictions consistent with free models, different from the finite-size exact diagonalization results discussed in the following. For simplicity, we calculate the level-spacing ratio and the entropy for the system with open boundary conditions. In the level-spacing ratio calculations, we consider appropriate sectors of the lattice reversal symmetry and the longitudinal and transverse parities.

The average half-chain entanglement entropy of typical fermionic Gaussian states is given by $S_{L/2}^{\rm gauss}=0.279$~\cite{lydzba2020eigenstate}. Since our mean-field eigenstates are Gaussian, it is not surprising that we find agreement with the above formula for any set of model parameters $g$ and $\eta$. On the contrary, we observe a non-trivial $\eta$ dependence of the mean eigenstate entropy by using finite-size exact diagonalization. Close to the boundary interaction angles $\eta=\pm\pi/4$, the mean entropy is smallest (though still larger as the Gaussian result) and then increases towards the middle (see Appendix~\ref{app: entropy} for details). The difference in entanglement scaling of the mean-field approximation and the exact finite-size results displays the limits of the Gaussian approximation, which correctly captures local observables (e.g., the spectrum shown in \fref{fig: ez specter}) but fails to capture global properties (e.g., entanglement entropy).

\begin{figure}[!htb]
   \includegraphics[width=1.0\columnwidth]{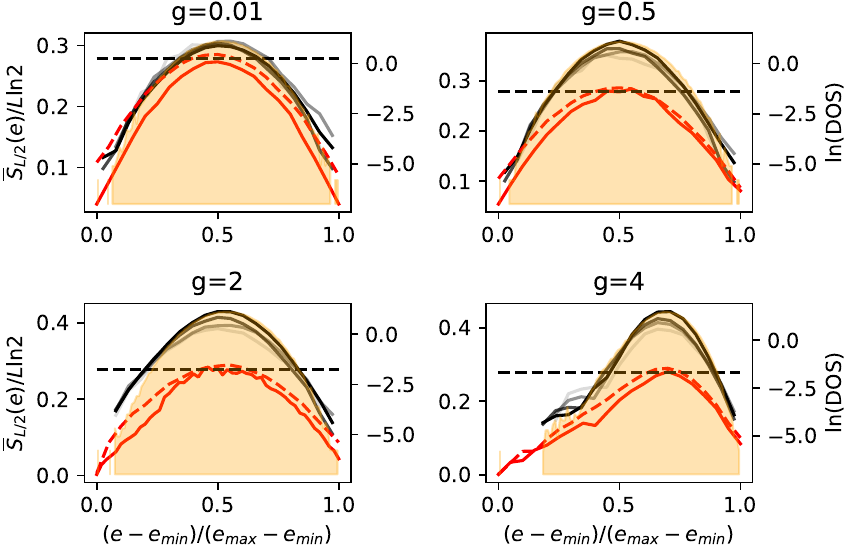}
    \caption{
    Half-chain entanglement entropy scaling at different energy densities. \textit{Left axes:} The full lines represent the mean half-chain entanglement entropy $\overline{S}_{L/2}$ for system sizes 8, 10, 12, 14, 16, 18 (from bright to dark). The dashed line is the Gaussian result $S_{L/2}^{\rm gauss}=0.279$. The red lines represent the mean entanglement entropy for the self-consistent eigenstates ($L=60$: full line, $L=16$: dashed line). \textit{Right axes:} The orange histogram shows the log of the density of states for the system size $L=18$.  We fix the interaction angle $\eta=0.15\pi$.  We perform the finite size calculations on the model with open boundary conditions.
}
\label{fig: ent scaling energy}
\end{figure}

Although the half-chain entanglement entropy average is larger than the typical Gaussian result, the states at the spectral boundaries should have a significantly lower entropy. We show this in \fref{fig: ent scaling energy} where we fix the interaction angle $\eta=0.15\pi$ and plot the scaling of the half-chain entanglement entropy at different energy densities. The mean entropy essentially determines the maximum entropy at maximum energy density (shown as a histogram in \fref{fig: ent scaling energy}). Therefore, the entanglement entropy scaling at large energy density matches the mean entanglement entropy scaling. Interestingly, the entanglement entropy shape matches the shape of the logarithm of the density of states and decreases towards the boundaries of the spectra. The entanglement entropy close to the spectral edges is thus near the mean-field result shown in red in \fref{fig: ent scaling energy}.

The discrepancy between the Gaussian result obtained from the mean-field theory and the exact finite-size entanglement entropy calculations indicates that the states in the middle of the spectrum might still be chaotic. 

\begin{figure}[!htb]
   \includegraphics[width=1.0\columnwidth]{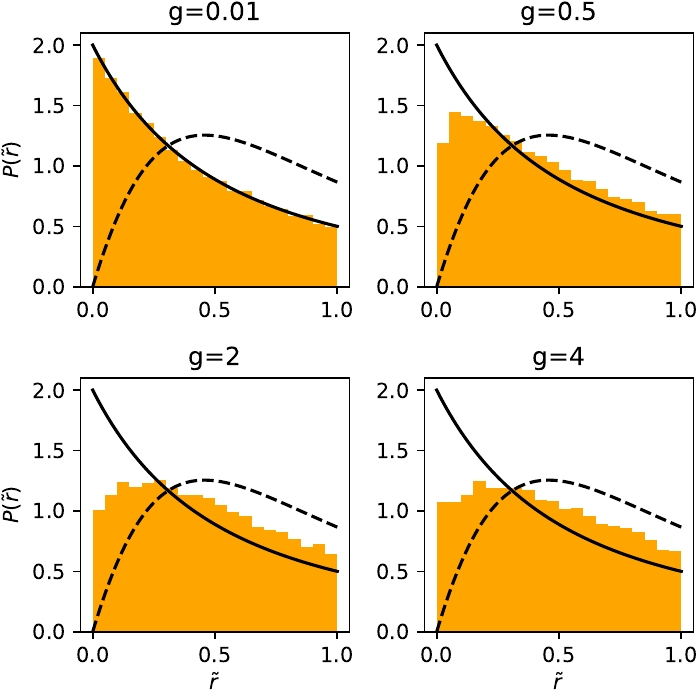}
    \caption{
    We obtain the distribution of level spacing ratios by exact diagonalization ($L=18$). At small $g$, it is close to the analytic result for the integrable case $P(\tilde{r})=\frac{1}{(1+\tilde{r})^2}$ (black line) \cite{atas2013distribution}. Upon increasing $g$, we obtain a distribution between the GOE (dashed line) and Poisson (black line) results. We perform the finite size calculations on the model with open boundary conditions.
}
\label{fig: level spacing}
\end{figure}

Therefore, we compare the distributions of level spacing ratios obtained from exact diagonalization to the Poisson and GOE values for integrable and quantum chaotic systems, respectively. In \fref{fig: level spacing}, we show that the distribution of level spacing ratios matches the integrable case at small interaction strengths. We see deviations from the Poisson distribution towards the GOE ensemble result upon increasing the infinite-range interaction, which matches the behavior of the entanglement entropy and shows that our mean-field treatment can not capture global eigenstate properties.
\section{Conclusions}
\label{sec: conclusion}
We studied a model with competing nearest-neighbor and infinite-range interactions in the mean-field limit. Since the resulting mean-field Hamiltonian is the (self-consistent) transverse-field XY model, we obtained several analytic zero-temperature results, including a non-analytic transition at the interaction angle $\eta=-\pi/4$. We also numerically determined the finite-temperature phase diagram and found a reentrant phase transition. Besides, we investigate the mean-field eigenstate and spectral properties of the model. We found that the mean-field approximation captures local observables as the magnetization and energy. However, the mean-field treatment fails to capture the entanglement entropy and level-spacing ratio distributions. Since the presented model is the simplest model displaying short- and long-range dichotomy, it would be interesting to extend our study of equilibrium properties of the model to the non-equilibrium case, where a dynamical phase transition  \cite{vzunkovivc2016dynamical, Zunkovic:2018aa,vzunkovivc2018dynamical} and chaotic dynamical phases~\cite{lerose2018chaotic,lerose2019impact} might be observed and studied analytically. Our approach can be extended by including quantum corrections and studying their finite-size scaling, which could provide more insight into the validity of the mean-field treatment.  

\section*{ACKNOWLEDGMENTS}
\begin{acknowledgments}
BZ acknowledges support from ARRS project J1-2480 and partial support from the European Research Council (ERC) under Advanced Grant No. 694544 -- OMNES. PR acknowledges partial support from Fundação para a Ciência e Tecnologia (Portugal) through Grant and UID/CTM/04540/2019. The computational resources were provided by the European Research Council (ERC) under Advanced Grant No. 694544
-- OMNES and the Slovenian national supercomputing network -- SLING.
\end{acknowledgments}

\bibliographystyle{apsrev4-1}
\bibliography{XY_Chain}

\newpage 

\appendix

\onecolumngrid

\section{Equilibrium Phase diagram \label{sec: Further-analysis-of-Phase Diagram}}
In this appendix, we present additional results regarding the equilibrium phase diagram. We first discuss the zero temperature and then the finite temperature case.

\subsection{Zero temperature}
At zero temperature the free energy density reduces to $f(\eta,g,\phi_{z})=\frac{gz^{2}}{2}-\frac{1}{\pi}\int_{0}^{\pi}\omega(q,\eta,g\phi_{z})\mathrm{d}q$.
The stability of the paramagnetic solution is determined by looking
at the second derivative of the free energy density concerning
the order parameter $\phi_{z}$ at $\phi_{z}=0$. We find the critical
interaction $g$ by first evaluating the second derivative of the
    dispersion relation with respect to the order parameter at $\phi_{z}=0$ and then integrating out the momentum. The resulting expression is given in terms of complete elliptic integrals of the first and the second kind
    \begin{align}
    \tilde{g}(\eta)/J & =\frac{\pi\sin(2\eta)}{\sqrt{1-\sin(2\eta)}\left[E(\tilde{\eta})-K(\tilde{\eta})\right]},\label{eq: unstable g}\\
    \tilde{\eta} & =2+\frac{2}{\sin(2\eta)-1}\nonumber 
    \end{align}
We obtain the critical field at which the paramagnetic solution $\phi_{z}=0$ becomes unstable by solving the remaining equation.

The boundaries of the interaction angle $\eta\in[-\pi/4,\pi/4]$ represent models with the anisotropic and the isotropic interaction. In these cases, we find the exact critical interaction strengths $g^*$ and the order parameter close to the phase transition.
\subsubsection{Anisotropic case: $\eta=-\pi/4$.}
At $\eta=-\pi/4$, the ordered phase persists for all $g>0$, vanishing
continuously at $g=0$. By applying perturbation
theory, we calculate the order parameter as a function of
the infinite-range interaction strength $g$. We calculate the free energy to second order in the order parameter $\phi_{z}$
\begin{equation}
f(\eta=-\pi/4,g,\phi_{z})=\frac{1}{2}g\left(\phi_{z}^{2}-\frac{4\left|\phi_{z}\right|E\left(-\frac{2}{g^{2}\phi_{z}^{2}}\right)}{\pi}\right)
\end{equation}
resulting in the self-consistency equation 
\begin{equation}
\phi_{z}=\frac{2K\left(-\frac{2}{g^{2}\phi_{z}^{2}}\right)}{\pi}.
\end{equation}
Expanding this relation up to the second order in $g$ yields
to the result reported in the main text, i.e., Eq.\eqref{eq: 3quat z-1}
\begin{equation}
\phi_{z}\left(g,\eta=-\frac{\pi}{4}\right)=\frac{4\sqrt{2}e^{-\frac{\pi}{\sqrt{2}g}}}{g}.\label{eq: app 3quat z-1}
\end{equation}

In the vicinity of $\eta=-\pi/4$, we perturbatively calculate the critical line close to the non-algebraic fixed point. To achieve this, we apply a perturbative expansion of the free energy in $\delta\eta=|\eta+\pi/4|$ and $g$
\begin{align*}
f(\delta\eta,g,\phi_{z}) & =-\frac{K\left(-\frac{2}{g^{2}\phi_{z}^{2}}\right)\left(\delta\eta^{2}+2\left(\delta\eta^{2}-1\right)g^{2}\phi_{z}^{2}\right)}{g\phi_{z}\pi}\\
 & +\frac{\left(\delta\eta^{2}\left(g^{2}\phi_{z}^{2}+3\right)-2\right)E\left(\frac{2}{g^{2}\phi_{z}^{2}+2}\right)-K\left(\frac{2}{g^{2}\phi_{z}^{2}+2}\right)\left(\delta\eta^{2}+\left(\delta\eta^{2}+2\right)g^{2}\phi_{z}^{2}\right)}{\pi\sqrt{g^{2}\phi_{z}^{2}+2}}\\
 & +\frac{g\phi_{z}\left(3\delta\eta^{2}+2\left(\delta\eta^{2}-1\right)g^{2}\phi_{z}^{2}-2\right)E\left(-\frac{2}{g^{2}\phi_{z}^{2}}\right)}{\pi g^{2}\phi_{z}^{2}+2}+\frac{g\phi_{z}^{2}}{2}.
\end{align*}
By expanding around $\eta=-\pi/4$ and using the ansatz
\begin{equation}
\eta=-\pi/4+4e^{-\left(1+\frac{\pi}{\sqrt{2}g^{*}\left(\eta\right)}\right)}
\label{eq: app critical angle}
\end{equation}
for the angle, we calculate the non-trivial free energy minimum close to the non-algebraic critical line

\begin{equation}
\phi_{z}(g,\eta=-\pi/4+\delta\eta)=\frac{4\sqrt{2}}{g}e^{\frac{1}{2}\left(W\left(-\frac{\epsilon^{2}}{e}\right)-\frac{\sqrt{2}\pi}{g}\right)},\label{eq: 3quat z-1-1}
\end{equation}

\begin{align*}
\delta\eta=4e^{-\frac{\pi}{\sqrt{2}g}-1/2}\epsilon,
\end{align*}
where $W(x)$ denotes the principal solution of the equation $x=W\exp W$.

\subsubsection{Isotropic case: $\eta=\pi/4$ }
At the right boundary, $\eta=\pi/4$, the model reduces to an XX model
with a self-consistent transverse field. The equations are simplified
since the XX hopping conserves the self-consistent transverse magnetization.
The free energy reduces to {\small{}
\begin{align}
f( & \eta=\pi/4,g,\phi_{z})=\frac{g\phi_{z}^{2}}{2}\\
 & -\frac{1}{\pi}\begin{cases}
\begin{array}{cc}
g\pi\phi_{z} & \left(g=\sqrt{2}\land\phi_{z}=1\right)\lor\left(g>\sqrt{2}\land g\phi_{z}\geq\sqrt{2}\right)\\
2\left(\sqrt{2}\cosh\left(\frac{\log(2)}{2}-\log\left(\sqrt{2-g^{2}\phi_{z}^{2}}-ig\phi_{z}\right)\right)+g\sin^{-1}\left(\frac{g\phi_{z}}{\sqrt{2}}\right)\phi_{z}\right) & \text{Otherwise}
\end{array}\end{cases}
\end{align}
} and the self-consistency condition to 
\begin{equation}
\pi g\phi_{z}=\begin{cases}
\begin{array}{cc}
\pi g & \left(g=\sqrt{2}\land\phi_{z}=1\right)\lor\left(g>\sqrt{2}\land g\phi_{z}\geq\sqrt{2}\right)\\
2g\sin^{-1}\left(\frac{g\phi_{z}}{\sqrt{2}}\right) & \text{Otherwise}
\end{array}\end{cases}.
\end{equation}
The minimum of the free energy density can thus be obtained only at
$\phi_{z}=0$ or $\phi_{z}=1$. By comparing the values of the free
the energy at those two values of the self-consistent field we get the
equations for the critical $g$

\begin{equation}
\pi g^{*}+4\sqrt{2}=\pi g^{*}
\end{equation}
with the solution reported in the main text. The order parameter vanishes below the critical interaction strength $g^{*}$ and is equal to one above the critical field.

\subsection{Finite temperature}

In this section, we discuss some perturbative results at finite temperatures. 

\subsubsection{Large infinite-range coupling}

First, we calculate the properties of the second-order transition at
the large infinite-range coupling. To this end, we expand the free energy
up to the second order in $J/g$

\begin{align*}
f(\eta,\phi_{z},g,T)=\frac{\phi_{z}^{2}}{2}-\frac{\left(T(1-\sin(2\eta))\tanh\left(\frac{\phi_{z}}{T}\right)+\phi_{z}(\sin(\eta)+\cos(\eta))^{2}\text{sech}^{2}\left(\frac{\phi_{z}}{T}\right)\right)}{4g^{2}T\phi_{z}}+T\log\left(\cosh\left(\frac{\phi_{z}}{T}\right)\right).
\end{align*}
Using the above expression in the stationarity equation, we obtain
a non-trivial stationary point 
\begin{align*}
\phi_{z}=\frac{T\sqrt{-3g^{2}(T-1)T^{2}-\sin(2\eta)-2}}{\sqrt{g^{2}T^{2}-\frac{8}{5}\sin(2\eta)-\frac{12}{5}}}.
\end{align*}
We observe that that the critical point is at $T=1+O((J/g)^{2})$
and that the transition is of mean-field type 
\begin{align*}
z=\sqrt{3(1-T)}.
\end{align*}

\subsubsection{Ising model $\eta=0$}

In the Ising case $\eta=0$, the free energy expanded up to fourth
order in $\phi_{z}$ simplifies to 
\begin{align*}
f(\eta,\phi_{z},g,T) & =\frac{3g^{4}\text{sech}^{4}\left(\frac{1}{T}\right)\phi_{z}^{4}}{32T^{3}}-\frac{g^{4}\text{sech}^{2}\left(\frac{1}{T}\right)\phi_{z}^{4}}{16T^{3}}\\
 & +\frac{g^{4}\tanh\left(\frac{1}{T}\right)\text{sech}^{2}\left(\frac{1}{T}\right)\phi_{z}^{4}}{16T^{2}}-\frac{1}{64}g^{4}\tanh\left(\frac{1}{T}\right)\phi_{z}^{4}+\frac{g^{4}\text{sech}^{2}\left(\frac{1}{T}\right)\phi_{z}^{4}}{64T}\\
 & -\frac{1}{4}g^{2}\tanh\left(\frac{1}{T}\right)\phi_{z}^{2}-\frac{g^{2}\text{sech}^{2}\left(\frac{1}{T}\right)\phi_{z}^{2}}{4T}+\frac{g\phi_{z}^{2}}{2}-T\log\left(2\cosh\left(\frac{1}{T}\right)\right).
\end{align*}
Inserting the above equation into the self-consistency equation, we obtain a non-trivial order parameter
\begin{align*}
\phi_{z}=\frac{2\sqrt{2}T\sqrt{T\left(g\tanh\left(\frac{1}{T}\right)-2\right)+g\text{sech}^{2}\left(\frac{1}{T}\right)}}{\sqrt{g^{3}\left(-T^{3}\tanh\left(\frac{1}{T}\right)+\left(T^{2}+4T\tanh\left(\frac{1}{T}\right)-4\right)\text{sech}^{2}\left(\frac{1}{T}\right)+6\text{sech}^{4}\left(\frac{1}{T}\right)\right)}}.
\end{align*}
The second-order critical interaction strength is then 
\begin{equation}
g_{\eta=0}^{*}/J=\frac{2T}{T\tanh\left(\frac{J}{T}\right)+J\text{sech}^{2}\left(\frac{J}{T}\right)}.\label{eq: g ising-1}
\end{equation}
Close to the second order transition, the order parameter is given
by 
\begin{align}
\phi_{z} & (\eta=0,\frac{T}{J},\frac{g_{\eta=0}^{*}}{J}+\frac{\delta g}{J})=\\
 & \frac{\sqrt{\frac{\text{\ensuremath{\delta}g}}{J}}\left(\tau+\text{sech}^{2}\left(\frac{J}{T}\right)\right)^{2}}{\sqrt{\frac{T}{J}\left(-\tau^{3}+\left(\frac{T^{2}}{J^{2}}+4\tau-4\right)\text{sech}^{2}\left(\frac{J}{T}\right)+6\text{sech}^{4}\left(\frac{J}{T}\right)\right)}}\nonumber \\
\tau & =\frac{T\tanh\left(\frac{J}{T}\right)}{J}.\nonumber 
\end{align}
The second-order transition is a simple mean-field transition. We also obtain the tricritical point in the Ising model by solving the following transcendental equation 
\begin{equation}
\frac{T^{3}\sinh\left(\frac{2J}{T}\right)}{2J^{3}}+4=\frac{T\left(\frac{T}{J}+4\tanh\left(\frac{J}{T}\right)\right)}{J}+6\text{sech}^{2}\left(\frac{J}{T}\right).\label{eq: tricritical ising-1}
\end{equation}
We checked numerically that the critical exponent $\nu$ remains mean-field-like up to the tricritical point for any interaction angle $\eta$.
At the tricritical point, the exponent retains its mean-field nature
and is $\nu=1/4$.

\section{Microcanonical ensemble}
In this appendix, we discuss the microcanonical properties of the XY model in a self-consistent transverse field.
\subsection{Multiple solutions of the self-consistency equation}
\begin{figure}[!h]
\includegraphics[width=0.45\textwidth]{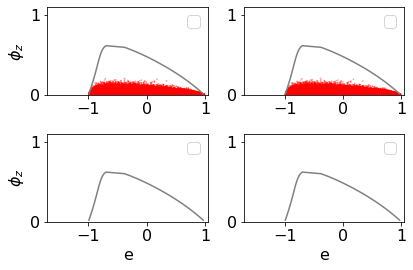}
\includegraphics[width=0.45\textwidth]{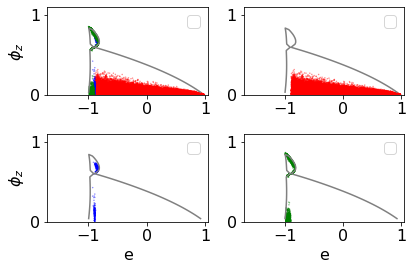}
\includegraphics[width=0.45\textwidth]{nsol_n200_g3.0_eta0.png}
\caption{
    Number of solutions for the one configuration of excitations: red (one solution), blue (three solutions), green (five solutions). 
 (\textit{a - Paramagnet}) parameters: $g=1.,\eta=0$. (\textit{b - reentrant phase transition}) parameters: $g=1.67,\eta=0$.(\textit{c - second order phase transition}) parameters: $g=3,\eta=0$.
}
\label{fig: no sol}
\end{figure}

The self-consistency equations for eigenstates are non-linear for each excitation configuration. We numerically observed that there is always at least one solution, but there also can be three or five solutions for one excitation configuration. Therefore, the number of solutions is larger than the Hilbert space. Although this is not surprising due to the non-linear nature of the equations, it raises the question of whether all states are physically relevant. We argue that in the thermodynamic limit, only one of the degenerate solutions becomes relevant. First, we observe that different self-consistent solutions for the same configuration of the excitations have a very different energy. Therefore, states in a narrow energy band that are part of distinct branches of self-consistent solutions, will have an exponentially different entropic factor. Therefore, in the thermodynamic limit, only one of the solutions remains relevant. In figure \fref{fig: no sol}, we show the degeneracy of the solutions for calculated eigenvectors. The degeneracy is related to the appearance of a new structure in the spectrum and is relevant for the first-order quantum and reentrant phase transitions.

\newpage
\subsection{System size dependence of the microcanonical order parameter \label{app: microcanonical convergence}}
\begin{figure}[!h]
\includegraphics[width=0.95\columnwidth]{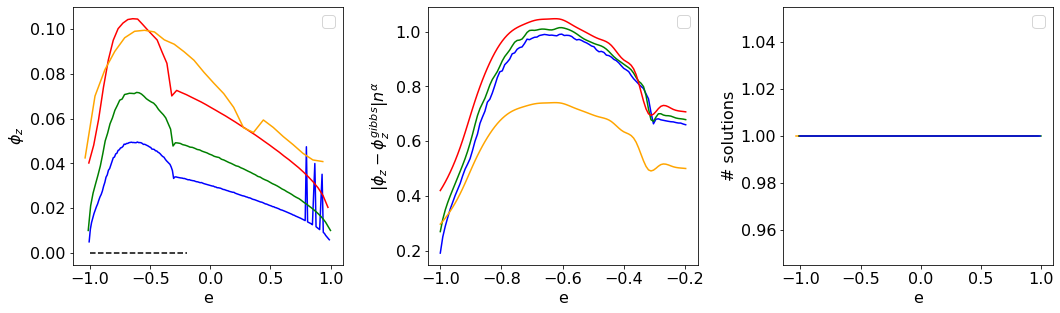}
\includegraphics[width=0.95\columnwidth]{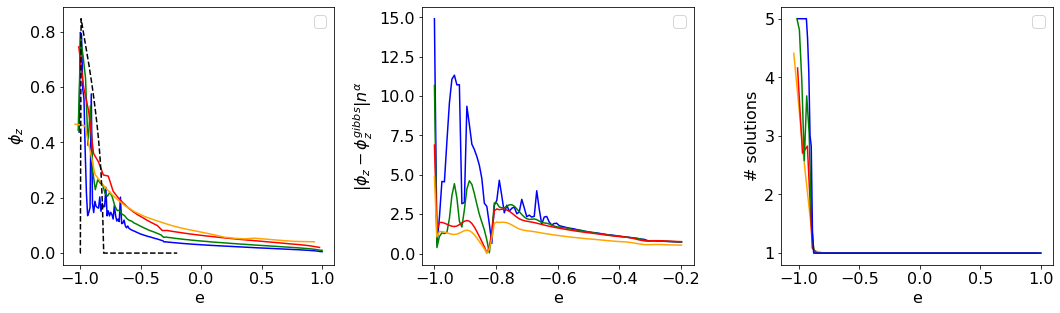}
\includegraphics[width=0.95\columnwidth]{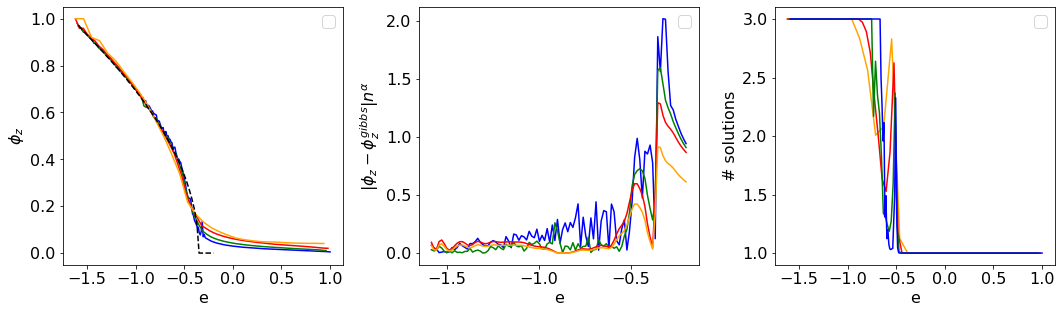}
\caption{
    Scaling of the microcanonical solution with the system size. In all plots, the scaling exponent is $\alpha=1/2$. (\textit{Top- Paramagnet}) parameters: $g=1.,\eta=0$. (\textit{Middle - reentrant phase transition}) parameters: $g=1.67,\eta=0$.(\textit{Bottom - second order phase transition}) parameters: $g=3,\eta=0$.
}
\label{fig: scaling}
\end{figure}

To check the microcanonical ensemble convergence to the Gibbs ensemble calculated in the previous section, we first numerically study the order parameter convergence with the system size. In \fref{fig: scaling}, we show the scaling of the order parameter in all regimes.

\newpage
\subsection{Convergence of the maximal eigenstate magnetization}
\begin{figure}[!htb]
    \includegraphics[width=0.45\textwidth]{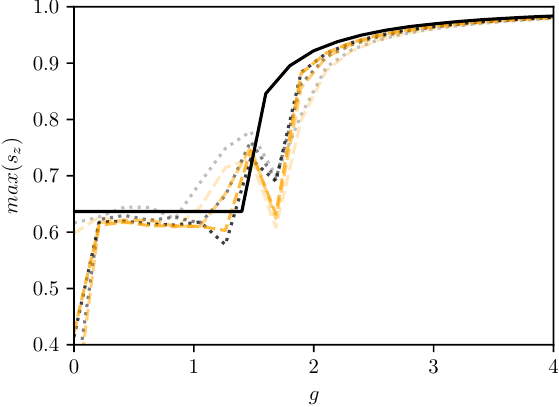}
    \caption{
        Scaling of the maximum eigenstate magnetization $\phi_{max}^z(g)$ with the system size: $L=$11, 15, 19 (from light to dark orange, dashed), 12, 16, 20 (from light to dark gray, dotted), and  $\infty$ (black), $\eta=0$. The finite-size calculations correspond to the XY chain with an infinite-range ZZ interaction.}
\label{fig: scaling_maxz}
\end{figure}

To calculate the spectral boundaries, we used the expression for the maximal magnetization of an eigenstate. We demonstrate how this maximal eigenstate magnetization converges with the system size. In \fref{fig: scaling_maxz}, we compare the exact diagonalization results for the Ising model with an infinite-range interaction for different system sizes. We observe that the convergence is very fast at large $g$ but is still far from the thermodynamic result close to the non-analytic point.

\subsection{Calculation of spectral boundaries}
\label{app: spectral boundaries}
We obtain spectral boundaries by finding the allowed magnetization range and then calculating the minimum and maximum energies for a given allowed magnetization value. The allowed magnetizations are obtained by calculating the maximum magnetization at a transverse magnetization $h$
\begin{align}
    s^{\rm z}_{\rm max}(h)=\frac{1}{\pi}\int_{0}^{\pi}{\rm d}q |s^{\rm z}_{q}(h)|,\quad s^{\rm z}_q(h)=\frac{h+\cos (q) (\sin (\eta )+\cos (\eta ))}{\sqrt{h^2+2 h \cos (q) (\sin (\eta )+\cos (\eta ))+\sin (2 \eta ) \cos (2 q)+1}}.
\end{align}
All magnetizations below the maximum magnetization $s^{\rm z}_{\rm max}$ that satisfy the self-consistency condition $h = gs^{\rm z}$ are allowed. We then equidistantly discretize the range of allowed eigenstate magnetizations and calculate the maximum and the minimum energy for each magnetization value. 

\begin{figure}[!htb]
   \includegraphics[width=0.3\columnwidth]{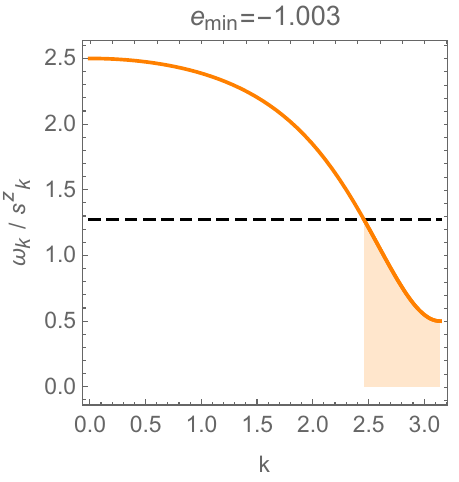}
   \includegraphics[width=0.3\columnwidth]{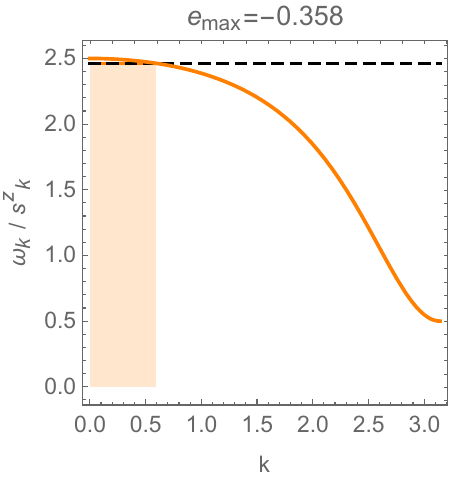}
    \caption{
    The figure displays the calculation of spectral boundaries at $\eta=0$, $g=3$ and $s^{\rm z}=0.5$. The orange line corresponds to the energy/magnetization ratio, and the black-dashed line denotes the ratio until we have to sum to obtain the correct magnetization. The left panel displays the minimum energy state, and the right panel the maximum energy state at a given magnetization. We obtain the corresponding state by exciting quasiparticles in the orange-shaded region of $k$.
}
\label{fig: spectral boundaries}
\end{figure}

The minimum and the maximum energy at a given magnetization are calculated by first calculating the ground state magnetization (and energy) and then adding excitations (see \fref{fig: spectral boundaries}). When calculating the minimum energy we add excitations that have the smallest ratio $\omega_{q}(h)/s^{\rm z}_q(h)$ until the eigenstate has the desired magnetization (\fref{fig: spectral boundaries} left). On the other hand, if we aim to calculate the maximum energy we add excitations that have the largest ratio $\omega_{q}(h)/s^{\rm z}_q(h)$ (\fref{fig: spectral boundaries} right).

\subsection{Entanglement entropy}
\label{app: entropy}
In this section, we first describe the parameter dependence of the mean entanglement entropy and then observe how the entanglement entropy depends on the order parameter if we fix the energy.

Let us first discuss the $\eta$ dependence of the mean entanglement entropy. At large $g>1$, the mean entanglement entropy increases with the system size for any interaction angle $\eta$. The increase is slower at the boundaries. Upon decreasing the infinite interaction strength, $g$, the increase of the entanglement entropy with the system size slows down. We recover the Gaussian result at small interaction strengths. 
\begin{figure}[!htb]
   \includegraphics[width=0.5\columnwidth]{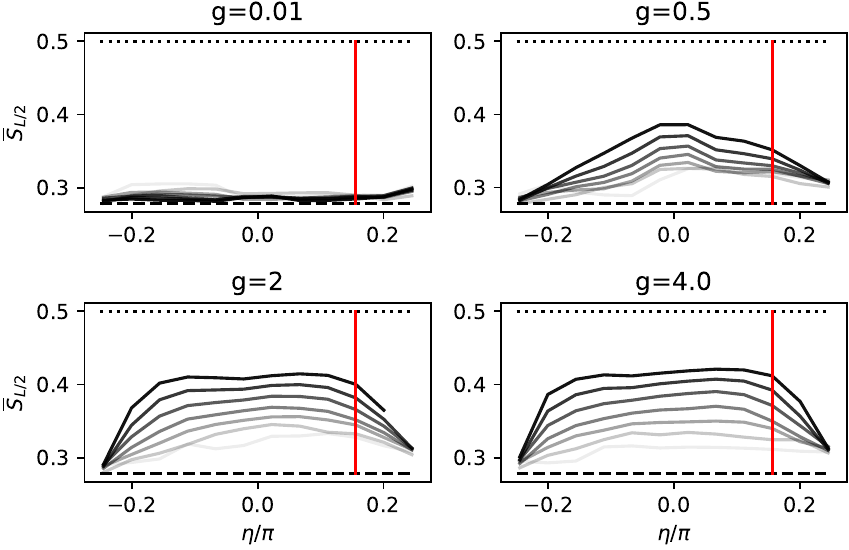}
    \caption{
    Half-chain entanglement entropy scaling. The full lines represent the mean half-chain entanglement entropy $\overline{S}_{L/2}$ for system sizes 6, 8, 10, 12, 14, 16, and 18 (from bright to dark). The dashed line is the Gaussian result ($S_{L/2}^{\rm gauss}=0.279$~\cite{lydzba2020eigenstate}), and the dotted line is the maximum volume-law entropy. The red line denotes the interaction angle for which we show the energy-dependent mean-entropy $\overline{S}_{L/2}(e)$ in \fref{fig: ent scaling energy}. We perform the finite size calculations on the model with open boundary conditions.
}
\label{fig: ent scaling}
f\end{figure}

\begin{figure}[!htb]
   \includegraphics[width=0.5\columnwidth]{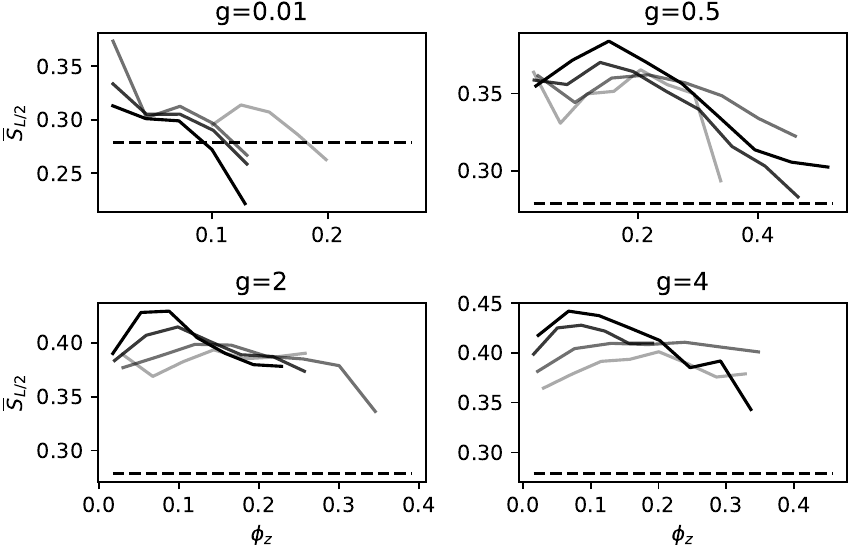}
    \caption{
    Half-chain entanglement entropy scaling at different order parameter $\phi_z$ for states close to the maximum energy density. The full lines represent the mean half-chain entanglement entropy $\overline{S}_{L/2}$ for system sizes 10, 12, 14, 16, and 18 (from bright to dark). The dashed line is the Gaussian result, $S_{L/2}^{\rm gauss}=0.279$~\cite{lydzba2020eigenstate}. We fix the interaction angle $\eta=0.15\pi$.
}
\label{fig: ent scaling mz}
\end{figure}

Finally, we checked how the mean entropy depends on the order parameter if we fix a small energy window close to the maximum energy density where the mean entropy is the largest. We find (see \fref{fig: ent scaling mz}) that at the boundaries of the spectrum, the mean entropy decreases with increasing system size. We also observe the decrease of the mean entropy close to the spectral edges at large $g$, where we find an increase in the mean entanglement entropy.

\end{document}